\documentclass[twocolumn,amsmath,amssymb,aps]{revtex4}
\usepackage{graphicx}
\usepackage{bm}
\newcommand{\be}{\begin{equation}}
\newcommand{\ee}{\end{equation}}
\newcommand{\beq}{\begin{eqnarray}}
\newcommand{\eeq}{\end{eqnarray}}

\begin{document}

\title{DNA unzipping and the unbinding of directed polymers in a random media.}

\author{Yariv Kafri$^1$ and Anatoli Polkovnikov$^{2}$}

\affiliation{$^1$ Department of Physics, Technion, Haifa 32000, Israel.\\
  $^2$Department of Physics, Boston University, Boston, MA 02215 }

\date{\today}

\begin{abstract}

We consider the unbinding of a directed polymer in a random media
from a wall in $d=1+1$ dimensions and a simple one-dimensional
model for DNA unzipping. Using the replica trick we show that the
restricted partition functions of these problems are {\em
identical} up to an overall normalization factor. Our finding
gives an  example of a generalization of the stochastic matrix
form decomposition to disordered systems; a method which
effectively allows to reduce dimensionality of the problem. The
equivalence between the two problems, for example, allows us to
derive the probability distribution for finding the directed
polymer a distance $z$ from the wall. We discuss implications of
these results for the related Kardar-Parisi-Zhang equation and the
asymmetric exclusion process.
\end{abstract}

\maketitle

The problem of a directed polymer in a random media (DPRM) has
received much attention for more than two decades \cite{Rev}. This
stems from many reasons: It is a relevant model for a single vortex
in a disordered type-II superconductor \cite{Nat} and it is one of
the simplest examples of disordered systems for which, in some
cases, exact result can be obtained. Moreover, the model maps both
to the Kardar-Parisi-Zhang (KPZ) equation \cite{KPZ}, which is
perhaps the simplest non-linear stochastic growth equation, and to
the noisy Burger's equation \cite{FHH}. The latter, in one
dimension, describes the long-time and large length-scale behavior
of asymmetric exclusion processes (ASEP)~\cite{BurgersAs, Krug},
which have been studied as prototypes of non-equilibrium systems
\cite{Schutz,Derrida}. The relations between the different models
has been extremely fruitful: in some cases results which are hard to
derive in one model can be easily obtained using another.

\begin{figure}[ht]
\center
\includegraphics[width=7cm]{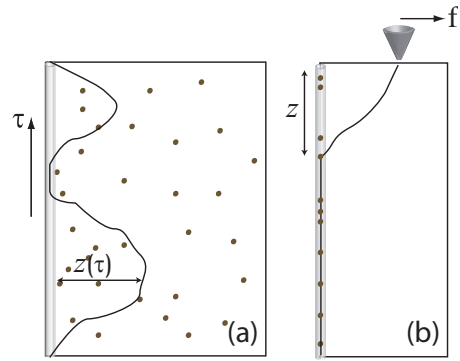}
\caption{The two problems discussed in the text. (a) Unbinding of a
line from an attractive wall in the presence of point disorder in
$d=1+1$. (b) Unzipping of a line with point disorder confined to an
attractive line. }\label{unzip}
\end{figure}

Here we present a new, more subtle, relation between the DPRM
problem and the DNA unzipping problem \cite{Lubensky}. In
particular, we consider (a) the depinning of a DPRM from an
attractive wall in $d=1+1$ dimensions and (b) the force induced
unzipping of a directed elastic line (in $d \geq 1+1$ dimensions)
from a disordered columnar defect. We study a low temperature
limit where the only excursion of the line from the defect occurs
at the edge of the sample where the external force is acting (see
Fig.~\ref{unzip}). Both problems are disordered. In the DPRM
problem the half-plane is taken to have uncorrelated point
disorder, while in the unzipping problem there is uncorrelated
point disorder localized on the columnar defect. In the DPRM
problem it is well known that as the strength of the disorder
grows there is an unbinding transition whereby the polymer leaves
the wall. Similarly, it is known that in the unzipping problem as
the force acting on the line exceeds a critical value, the line
leaves the potential. The unzipping model has been recently used
in the context of single molecule experiments performed on
DNA~\cite{Lubensky,Bhat,Prentiss} and also in relation to magnetic
force microscopy experiments in type-II
superconductors~\cite{knp}.

Both problems have been treated in some detail before and
superficially bear little resemblance. For example, the unbinding of
the DPRM is a continuous phase transition while the unzipping
problem is a first-order phase transition. More interesting, the
dimensionality of the problems is different. As we show the
unzipping problem is effectively one-dimensional, while the DPRM
problem is two dimensional. Nevertheless, we find that below the
unbinding/unzipping transitions the replicated partition function of
the DNA unzipping model is {\it identical} to the replicated
generating function for the localization length of the polymer near
the wall. Such an identity allows us to apply all known results from
the unzipping problem to the DPRM unbinding problem. For example, we
can derive the distribution function of the DPRM localization length
near the wall below the transition. This identity also allows one to
perform large scale numerical simulations of the DPRM by effective
reduction of the dimensionality of the problem.

We also discuss the implication of our result to both the KPZ
equation and the ASEP. When the DPRM model is mapped to the KPZ
equation, the wall acts as a free boundary condition for the
growing interface and the attractive potential reduces the growth
locally near the interface. The transition is approached by
increasing the local growth rate at the interface. The relation to
the unzipping problem allows us to derive the distribution
function for the height profile near the transition. When the DPRM
is mapped to an ASEP the boundary corresponds to particles being
injected at the end of a semi-infinite system with a rate which is
slower than the particle hopping rate in the bulk of the system.
Here the transition is approached by increasing the injection rate
of particles. Using our results we discuss the structure of the
particle density near the transition. The relation between the
models suggests a deeper relation which may not be transparent in
our derivation.

Equivalence between a two-dimensional (2d) unbinding problem and a
one-dimensional (1d) unzipping problem is quite remarkable because
lower dimensional problems are generally much easier to deal with.
Without disorder our results can be understood in the context of
stochastic matrix form (SMF) decomposition~\cite{chamon} as
follows. First, we map the partition function of a 2d classical
problem (DPRM) into the partition function of a quantum 1d
problem, where the coordinate along the polymer is replaced by an
imaginary time. Next we interpret the ground state wave-function
of this quantum problem as a partition function of another
classical 1d problem (DNA unzipping). The latter step is called
SMF decomposition. Here we present an example extending this
method to a disordered problem which corresponds to a Hamiltonian
which depends on (imaginary) time. We believe that our findings
can be further generalized to a wider class of disordered
problems.


{\em Unzipping of an elastic line from a columnar pin.} Let us
consider an external force ${\bf f}$ pulling the top end of a line
from an attractive disordered columnar pin (see Fig.~1b). Neglecting
excursions of the bound part of the line into the bulk, we can write
the partition function as a sum over positions $z$ where the line
leaves the pin. The contribution of the unzipped part can be easily
computed from the elastic energy~\cite{knp}:
\begin{equation}
{\cal F}_0(z)\!=-{\bf f}\cdot {\bf r(0)}+\!\!\!\int_0^z \!\!\!\!d
z' \left[ \frac{\gamma}{2} (\partial_{z' }{\bf r}(z'))^2\right] .
\label{f0}
\end{equation}
Here ${\bf r}(z)$ denotes transverse coordinates of the line and $z$
is the coordinate parallel to the defect. Integrating over all
possible paths gives the corresponding free energy: $F_0(z)=- f^2 z/
2\gamma$~\footnote{We note that there is a slight correction to this
expression, because the line, once it leaves the potential, is not
allowed to return. However, this correction will be negligible near
the unzipping transition, where $z$ is large.}. The free energy of
the bound part, up to a constant, is $F_1(z)=V_0(L-z)+\int_z^L d z'
U(z')$, where $V_0$ is the value of the attractive potential, $L$ is
the length of the columnar defect which is assumed to be very large,
and $U(z)$ is a random uncorrelated potential with zero mean
satisfying $\overline{U(z_1)U(z_2)}=\Delta\delta(z_1-z_2)$.

The partition function $Z$ is a sum of the corresponding Boltzmann
weights over all possible values of $z$~\cite{knp}:
\begin{equation}
Z=\int_0^L Z(z)=\int_0^L dz\, \mathrm e^{-F_{\rm uz}(z)},
\label{Z1}
\end{equation}
where up to an unimportant constant additive term $V_0 L$:
\begin{equation}
F_{\rm uz}(z)=F_0(z)+F_1(z)= \epsilon z+\int_z^L d z' U(z').
\label{fz}
\end{equation}
For simplicity we work in the units, where $k_B T=1$.  In the
equation above $\epsilon=(f_c^2-f^2)/2\gamma$, where $f$ is the
force applied to the end of the flux line and $f_c=\sqrt{2\gamma
|V_0|}$ is the critical force. Note that the partition function
(\ref{Z1}) describes effectively a one-dimensional problem.
The statistical properties of the unzipping transition can be
obtained by replicating Eq.~(\ref{Z1})~\cite{anderson}:
\begin{equation}
Z_n\equiv\overline{Z^n}=\overline{\int_0^L dz_1\ldots\int_0^L
dz_n\,\mathrm e^{-\sum_{\alpha=1}^n F_{\rm uz}(z_\alpha)}},
\label{Z_n}
\end{equation}
where the overline denotes averaging over point disorder. The
averaging procedure can be easily done for a positive integer $n$.
First we order the coordinates $z_j$, where the $j^{th}$ replica
unbinds from the pin according to: $0\leq z_1\leq
z_{2}\leq\dots\leq z_n$. Then for $z\in[0,z_1)$ there are no
replicas bound to the columnar pin, for $z \in[z_1,z_2)$ there is
one replica on the pin until finally for $L \geq z\geq z_n$ all
$n$ replicas are bound to the pin. Using this observation and
explicitly averaging over the point disorder in Eq.~(\ref{Z_n}) we
arrive at:
\begin{eqnarray}
Z_n=n!\int\dots\!\!\!\!\!\!\!\!\!\!\!\!\!\! \int\limits_{0\leq
z_1\dots\leq z_n\leq L}\!\!\!\!\!\!\!\! dz_n\dots dz_n\,\mathrm
e^{-\sum\limits_{j=1}^n \epsilon z_j- \Delta/2 j^2(z_{j+1}-z_{j})}
\label{tuam2}
\end{eqnarray}
where we use the convention $z_{n+1}=L$. The integral above is
straightforward to evaluate in the $L \to \infty$ limit:
\begin{eqnarray}
&&Z_n=e^{n^2\Delta/2\, L}{1\over \epsilon_n^n}\prod_{j=1}^n {1\over
1-\kappa_n j} \nonumber
\\
&&=e^{n^2\Delta/2\,
L}\left({2\over\Delta}\right)^n{\Gamma(1+1/\kappa_n-n)\over\Gamma(1+1/\kappa_n)}\equiv
e^{n^2\Delta/2\, L} Q_n.\phantom{XXX} \label{eq:partunzip}
\end{eqnarray}
where $\epsilon_n=\epsilon+\Delta n$ and
$\kappa_n=\Delta/2\epsilon_n$. The exponential prefactor is the
contribution of the whole pin while the rest of the expression is
the ($L$ independent) contribution from the unzipped region. The
disorder averaged free energy is given by the limit
$\overline{F}=-\lim_{n \to 0} (Z_n-1)/n$. Using
Eq.~(\ref{eq:partunzip}) one obtains
\begin{equation}
\overline{F}=\ln (\epsilon \kappa) + \Psi(1/\kappa),
\label{free_en}
\end{equation}
where $\Psi(x)$ is the digamma function and
$\kappa=\Delta/2\epsilon$. The unzipping transition occurs at
$\epsilon=0$ or equivalently at $\kappa \to \infty$. The expression
(\ref{free_en}) is identical to the one found in
Ref.~[\onlinecite{oper}] using the Fokker-Planck equation approach,
supporting the validity of the replica calculation.

{\em Unbinding from a wall.} Next we consider the unbinding of a
directed polymer (or interface) from an attractive wall, in the
presence of point disorder in the bulk of the system, in $d=1+1$
dimensions (see Fig.~\ref{unzip}a). The elastic free-energy of the
problem is given by
\begin{equation}
F_{\rm ub}=\int d\tau \left[\frac{\gamma}{2} (\partial_{\tau
}z(\tau))^2 +V(z(\tau)) + \mu(z(\tau),\tau) \right].
\end{equation}
Here $z(\tau)$ denotes the distance of the polymer from the wall at
position $\tau$, $\gamma$ is the line tension, $V(z)$ is a short
range attractive potential near the wall placed at $z=0$ and
$\mu(z,\tau)$ is the contribution from the point disorder. The free
energy of this problem was obtained first, using a replica
calculation, by Kardar~\cite{kardar}. For completeness, here we
outline the main points of the derivation. The overall weight of
paths connecting points $(0,0)$ and $(z,\tau)$, $W(z,\tau)$, can be
calculated from
\begin{eqnarray}
-\partial_\tau W(z,\tau) &=& \left[\mu(z,\tau)+V(z)- \gamma
\partial^2_z \right]W(z,\tau)\nonumber\\
&=&{\cal H}(z,\tau) W(z,\tau).\label{DP1}
\end{eqnarray}
After replicating the Hamiltonian $n$ times one obtains
\begin{equation}
{\cal H}_n= -{\sigma\over 2} n - \sum_{\alpha=1}^n \left[ \gamma
 \partial^2_{z_\alpha}\!+ V(z_\alpha) \right]\!
 -\sigma \sum_{\alpha < \beta} \delta(z_\alpha-z_\beta) \;,
\end{equation}
where we have assumed a Gaussian distribution of $\mu$ with zero
mean and variance $\sigma$. Averaging over disorder eliminated the
$z$ dependence of $\mu$. The ground-state wave function $\psi$ (as
well as the energy) of the problem were obtained by Kardar using the
Bethe ansatz. We skip the details as they can be found in
Ref.~\cite{kardar}. For the permutation ${\bf P}$ of particles such
that $0<z_{P1}<z_{P2}< \ldots <z_{Pn}$ and below the transition
$\psi \sim \exp \left(-\sum_{\alpha=1}^n \chi_\alpha z_{P
\alpha}\right) $. Here $\chi_\alpha = \lambda+2(\alpha-1)\chi$,
$\chi= \sigma/4 \gamma$ and $\lambda$ depends on the strength of the
attractive potential at the wall. We note that the wave function
gives the probability distribution of the directed polymer at the
top of the sample ($\tau=L$), and thus it is proportional to the
restricted partition function $W(L,z)$. Integrating $\psi$ over all
coordinates gives the generating function of moments of
$z$~\cite{kardar}:
\begin{equation}
N_n\equiv\int dz_1\dots\int dz_n\,
\psi={1\over\chi^n}{\Gamma(1+1/\kappa_n-n)\over\Gamma(1+1/\kappa_n)},
\end{equation}
where now $\kappa_n=\chi /\epsilon_n$ and
$\epsilon_n=\lambda-\chi+2\chi n$. Note that with a proper
identification of $\epsilon_n$ and $\kappa_n$
\begin{equation}
Q_n=N_n \;. \label{main}
\end{equation}
This is the main result of our paper. In fact, one can check that
not only the generating functions $Q_n$ and $N_n$ coincide, but also
the wave function of the DPRM problem is identical (up to an
unimportant constant) to the restricted partition function of the
DNA unzipping problem. This equivalence implies that {\it below the
transition} all the moments and cumulants of the unzipped length and
of the distance of the polymer from the wall are equivalent. In
particular, the weight of finding the DPRM a distance $z$ from the
wall for a {\it particular} realization of disorder is given by
$Z(z)$, defined in Eq.~(\ref{Z1}). Note that the free-energies of
the two models are {\it distinct}. Indeed $Q_n$ contains information
on the free-energy of the unzipping problem. However, for the DPRM
$N_n$ only describes the spatial distribution of $z$ at the top of
the sample. In particular, the sum over weights of a DPRM, which
ends at the point $z$ at the top of the sample, $W(z,L)$, is given
by
\begin{equation}
W(z,L)=e^{-E_0 L} \frac{\psi(z)}{\int dz' \psi(z')}=e^{-E_0 L}
\frac{Z(z)}{Z}, \label{rel1}
\end{equation}
where $Z(z)$ is defined in Eq.~(\ref{Z1}), $E_0$ is a sample
dependent free-energy and $Z$ is the normalization factor. We note
that the equivalence between $\psi(z)$ and $Z(z)$ gives an example
of the SMF decomposition~\cite{chamon} for a disordered DPRM
problem corresponding to a time-dependent Hamiltonian.

It is easy to see that for both models
\begin{equation}
\overline{\langle z\rangle}=\partial
F/\partial\epsilon=(\kappa\epsilon)^{-1}\Psi^{(1)}(1/\kappa),
\label{zav}
\end{equation}
where $\Psi^{(n)}(x)$ stands for the $n$-th derivative of the
digamma function. The expression above predicts a crossover from
$\overline{\langle z\rangle}\approx 1/\epsilon$ for $\kappa\ll 1$
(far from the transition) to $\overline{\langle
z\rangle}\approx\kappa/\epsilon=\Delta/\epsilon^2$ for $\kappa\gg 1$
(close to the transition) as was noted previously for both
unzipping~\cite{Lubensky} and unbinding~\cite{kardar} problems
separately. Similarly
\begin{equation}
w=\overline{\langle z^2 \rangle - \langle z
\rangle^2}=\partial^2F/\partial\epsilon^2=-(\kappa\epsilon)^{-2}\Psi^{(2)}(1/\kappa).
\label{fav}
\end{equation}
Here there is a crossover from $w \approx 1/\epsilon^2$ for $\kappa
\ll 1$ to $w \approx 2 \kappa/\epsilon^2=\Delta/\epsilon^3$ for
$\kappa\gg 1$. As has been noted in the context of DNA unzipping
$\sqrt{w}/\overline{\langle z\rangle}$ changes from being unity for
$\kappa \ll 1$ to $\sim \epsilon^{1/2}$ for $\kappa \gg 1$. Thus
close to the transition, thermal fluctuations become negligible.

{\em Calculation of the second moment.} With a little more work we
can evaluate higher moments of the distribution. In particular, the
second moment, which gives the variance. To do this we consider the
generating function:
\be
{\cal W}_n=n!\int dz_1\dots\!\!\!\!
\!\!\!\!\!\!\!\int\limits_{\!\!\!\!\!\!\!\!\!\!0\leq z_n\dots\leq
z_1\leq \infty}\!\!\!\!\!\!\!\!\!\!\! dz_n\,\mathrm
e^{-\sum\limits_{j=1}^n \epsilon_jz_j+\Delta/2
j^2(z_{j-1}-z_j)}\!\!.
\label{zm1}
\ee
The second (and similarly the higher) moments can be found by
differentiating ${\cal W}_n$ with respect to $\epsilon_j$:
\be
\overline{\langle z^2\rangle}=\lim_{n\to 0} \left. {1\over {\cal
W}_n}\,{1\over n}\sum_{j=1}^n {\partial^2 {\cal W}_n\over\partial
\epsilon_j^2}\right|_{\epsilon_j=\epsilon}.
\label{zm3}
\ee
Evaluating the simple integral in Eq.~(\ref{zm1}) we find that
${\cal W}_n=1/\prod_{j=1}^n \left[\sum_{k=1}^j\epsilon_k\,-\,\Delta
j^2/2\right]$ and correspondingly
\be
\overline {\langle z^2\rangle}={1\over
\epsilon^2}\lim_{n\to 0}{1\over n}\sum_{j=1}^n {2\over 1-\kappa
j}\sum_{k=j}^n {1\over k (1-\kappa k)}.
\ee
This sum can be evaluated using a trick similar to the one suggested
by Kardar~\cite{kardar}:
\beq
\overline{\langle z^2\rangle}&=&{2\kappa^2\over \epsilon
^2}\int\!\!\!\!\!\!\int\limits_{\!\!\!\!x>y>0}\!\!\!\! dx dy {1\over
\mathrm e^{\kappa x}-1}{y\,\mathrm e^{-y}\over \mathrm e^{\kappa y
}-1}\left[ \mathrm e^{\kappa y}+\mathrm e^{2y}\mathrm e^{\kappa
x-x}\right]\nonumber\\
&-&{4\over \kappa
\epsilon^2}\Psi^{(1)}(1/\kappa)\left(C+\Psi(1/\kappa)\right),
\label{z2} \eeq
where $C\approx 0.577$ is the Euler's constant. In the limit of weak
disorder or high temperature $\kappa\ll 1$, not surprisingly, we get
$\overline{\langle z^2\rangle }\approx 2/ \epsilon^2$, which agrees
with the Poissonian statistics of $z$ with a average given by
$\overline{\langle z \rangle}=1/\epsilon$. In the opposite limit
$\kappa\gg 1$ one finds $\overline{\langle z^2\rangle }=4\kappa^2/
\epsilon^2=4\overline{\langle z\rangle}^2$. Thus in this limit the
relative width of the distribution $\delta z/\overline{\langle
z\rangle}$, where $\delta z^2=\overline{\langle
z^2\rangle}-\overline{\langle z\rangle}\!\!\phantom{x}^{2}$, is
larger by a factor of $\sqrt{3}$ than that in the high temperature
regime. The distribution becomes superpoissonian at large $\kappa$.
In fact at $\kappa\to\infty$ one can derive the full distribution
function $P_{\kappa\to\infty}(z)$ using extreme value
statistics~\cite{Lubensky, ledoussal}: ${\cal
P}_{\kappa\to\infty}(z)\approx {\epsilon/ \kappa}\,
G(z\,\epsilon/\kappa)$, with
\be
G(x)={1\over\sqrt{\pi x}}\,\mathrm
e^{-x/4}-{1\over 2}{\rm erfc}(\sqrt{x}/2),
\label{dist}
\ee
where ${\rm erfc}(x)$ is the complimentary error function. It is
easy to check that this distribution indeed reproduces correct
expressions for the mean and the variance. We additionally performed
direct numerical simulations of the partition function (\ref{Z1})
and got excellent agreement with predictions of Eqs.~(\ref{z2}) and
(\ref{dist})~\cite{kp}.

\begin{figure}[h]
\center
\includegraphics[width=7cm]{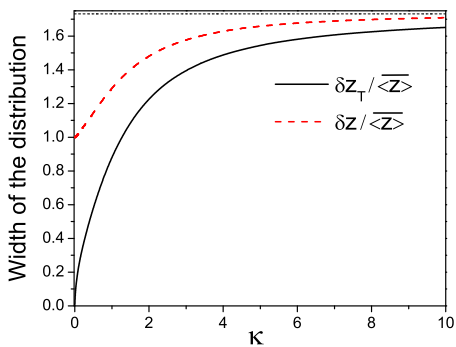}
\caption{Dependence of the relative width of the distribution on the
disorder parameter $\kappa$. The two curves correspond to different
averaging over temperature and disorder (see the text for details).
The horizontal line at $\sqrt{3}$ denotes the asymptotic value of
both $\delta z/\overline{\langle z\rangle}$ and $\delta
z_T/\overline{\langle z\rangle}$ at $\kappa\to\infty$}\label{fig_dz}
\end{figure}

The quantity $\overline{\langle z^2\rangle}$ is not always
experimentally accessible. For example, in the unzipping experiments
it is easier to measure thermal average, $\langle z\rangle$, in each
experimental run. Then the variance of the distribution will be
characterized by $\overline{\langle z\rangle^2}$. The difference
between the two expectation values is given by $w$ found in
Eq.~(\ref{fav}). Defining $\delta z_T^2=\overline{\langle
z\rangle^2}-\overline{\langle z\rangle}^{\,2}$ and using
Eqs.~(\ref{z2}) and (\ref{fav}) we find that $\delta
z_T/\overline{\langle z\rangle}\approx \sqrt{\kappa/2}$ in the weak
disorder limit ($\kappa\ll 1$) and $\delta z_T/\overline{\langle
z\rangle}\approx \sqrt{3}-1/(\sqrt{3}\kappa)$ for $\kappa\gg 1$. We
plot both $\delta z_T$ and $\delta z$ versus the disorder parameter
$\kappa$ in Fig.~\ref{fig_dz}.
Note that importance of the order of thermal and disorder averaging
also appears in the calculation of higher moments of $z$ becoming
irrelevant only in the zero-temperature (strong disorder) limit
($\kappa\to\infty$).

{\em Implications.} We now turn to discuss the implications of our
results for the KPZ equation and the ASEP. The height variable in
the KPZ equation is well known to be related to the DPRM through the
Cole-Hopf transformation $h(z,\tau)=-\ln(W(z,\tau))$. The latter
together with Eq.~(\ref{DP1}) yields
\begin{equation}
\partial_\tau h(z,\tau)= \gamma \partial_z^2 h(z,\tau) - \gamma
(\partial_z h(z,\tau))^2 + V(z)+\mu(z,\tau) \;. \label{KPZ}
\end{equation}
Now $\tau$ represents a time coordinate for the growing interface.
The pinning potential at the origin leads to a reduced growth rate
at a free boundary of the interface (for more details on the
correspondence see, for example,~\cite{Krug}). As stated previously,
the relation between the unzipping problem and the DPRM does not
give the full information about $W(z,\tau)$. It only contains its
spatial behavior but misses the prefactor (see Eq.~(\ref{rel1})),
which corresponds in the KPZ picture to the overall height of the
interface. However, the mapping implies that up to this overall
height the interface is described by Eq.~(\ref{fz}), namely
$h(z,\tau)=h_a(\tau)+F_{\rm uz}(z)$, where $h_a(\tau)$ is the mean
height and $F_{\rm uz}(z)$ is a tilted random walk. The probability
for an interface profile $F_{\rm uz}(z)$ is equal to the probability
of drawing the random contributions $U(z)$ (see Eq.~(\ref{fz})). The
correspondence implies, for example, that $h(z)-h(0)$ has a Gaussian
distribution with a variance which increases linearly with $z$. The
localization of the polymer near the unbinding transition
corresponds to a single dominant minima in the free energy
(equivalent to a minimum in the interface), the average distance of
these minima from the interface behaves as $1/\epsilon^2$ (see the
discussion after Eq.~(\ref{zav})).

Next, we turn to implications of our results for the ASEP. We
consider a semi-infinite system where particles hop into a one
dimensional lattice at the left end with a rate $\alpha$ and hop in
the bulk of the system to the nearest neighbor on the right with a
rate 1. In the continuum limit the behavior of the system is
captured by the equation \cite{BurgersAs}
\begin{equation}
\partial_\tau \rho(z)= -\frac{1}{2} \partial_z^2 \rho(z) + 2\rho(z)
\partial_z \rho(z)-\partial_z \rho(z) + U(x) +
\eta(z,\tau) \;. \label{ASEP}
\end{equation}
where $U(x)$ represents injection of particles into the system and
$\eta(z,\tau)=\partial_z \mu(z,\tau)$ (with $\mu(z,\tau)$ as
before drawn from a Gaussian distribution) is a conserving noise.
The equation can be obtained from Eq.~(\ref{KPZ}) by setting
$\gamma=1/2$ and defining a variable $m(z,\tau)=(z-h(z,\tau))/2$
(note that $\partial_z h(z,\tau)$ satisfies the noisy Burgers
equation). It is straightforward to verify that
$\rho(z,\tau)=\partial_z m(z,\tau)$ indeed satisfies
Eq.~(\ref{ASEP}). Using the same reasoning as in the KPZ equation
we find that in the steady state the density profile is described
by $ \rho(z)=(1-\epsilon-U(z))/2$. This result suggests that in a
discrete realization of the model a particle appears at a given
site with a constant probability and there are no correlations
between different sites. Our results hold, in ceratin limits, even
when another boundary is included in the system. For example when
at the other end the particles are ejected at a high rate. Indeed
it is known that for the ASEP with open boundaries the density is
flat near the end where particles are injected into the
system~\cite{Schutz,Krug}. However, we emphasize the {\it lack of
correlations} near the left boundary arbitrarily close to a
continuous transition. Similarly, for the interface growth with
two boundaries our results apply if the growth rate is
sufficiently slow only near one of the walls. We have performed
numerical simulation which verified these statements. These
results will be presented elsewhere~\cite{kp}.

In conclusion, we demonstrated that there is exact mapping between
the partition function of the DNA unzipping transition and the
spatial distribution of a DPRM unbinding from a wall. This mapping
allowed us to apply some known and newly derived results of the
simpler unzipping problem to the DPRM problem. We also showed how
this mapping can be used to derive results for the KPZ equation near
a boundary and about asymmetric exclusion process with open
boundaries.

We would like to acknowledge D.~R.~Nelson for useful discussions and
collaboration on a related work. We are also grateful for
illuminating discussions to C.~Castelnovo, C.~Chamon, and M.~R.
Evans. YK would like to thank the Newton Institute in Cambridge, UK
where part of the work was carried out. Work by Y. K. was supported
by the United States-Israel Binational Science Foundation (BSF)
through Grant. No. 2004072.


\begin{thebibliography}{99}

\bibitem{Rev} T. Halpin-Healy and Y. C. Zhang, Phys. Rep. {\bf
254}, 215 (1995).

\bibitem{Nat} C. A. Bolle et. al., Nature {\bf 399}, 43 (1999).

\bibitem{KPZ} M. Kardar. G. Parisi, Y. C. Zhang, Phys. Rev. Lett.
{\bf 56}, 889 (1986).

\bibitem{FHH} D. A. Huse, C. L. Henley and D. S. Fisher, Phys.
Rev. Lett. {\bf 55}, 2924 (1985).

\bibitem{BurgersAs} H. Van Beijeren, R. Kunter and H. Spohn, Phys.
Rev. Lett., {\bf 54}, 2026 (1985).

\bibitem{Krug} J. Krug and L-H Tang, Phys. Rev. E {\bf 50}, 104
(1994).

\bibitem{Schutz} For a review see, G. M. Schutz 2001 {\it
Physe Transitions and Critical Phenomena}, Vol. 19 ed C. Domb and
J. L. Lebowitz (New York, Academic).

\bibitem{Derrida} B. Derrida, Phys. Rep. {\bf 301}, 65 (1998).

\bibitem{Lubensky} D. K. Lubensky and D. R. Nelson, Phys. Rev. Lett.
{\bf 85}, 1572 (2000);  Phys. Rev. E {\bf 65}, 031917 (2002).

\bibitem{Bhat} S. M. Bhattacharjee, J. Phys. A {\bf 33}, L423
(2000).

\bibitem{Prentiss} C. Danilowicz, Y. Kafri, R. S. Conroy, V. W.
Coljee, J. Weeks and M. Prentiss, Phys. Rev. Lett. {\bf 93},
078101 (2004).

\bibitem{knp} Y.~Kafri, D.~R.~Nelson, and A.~Polkovnikov,
Europhys. Lett. {\bf 73}, 253 (2006).

\bibitem{chamon} C.~Castelnovo, C.~Chamon, C.~Mudry, P.~Pujol,
Annals of Phys. {\bf 318}, 316 (2005).

\bibitem{anderson} S.F.~Edwards and P.W.~Anderson, J. Phys. F {\bf
5}, 965 (1975).


\bibitem{oper} M. Opper, J. Phys. A {\bf 26}, L719 (1993).

\bibitem{kardar} M. Kardar, Nucl. Phys. B {\bf 290}, 582 (1987).

\bibitem{ledoussal} P. le Doussal, C.~Monthus, and D.~S.~Fisher,
Phys. Rev. E~{\bf 59}, 4795 (1999).

\bibitem{kp} Y.~Kafri and A.~Polkovnikov, to be published.

\end{thebibliography}
\end{document}